\documentclass[twocolumns,9pt]{iopart}
\usepackage{iopams}
\usepackage{graphicx,epsf}
\usepackage{color}
\usepackage{mathrsfs}
\usepackage{dcolumn}
\usepackage{cite}

\begin{document}
\title[]{Fine Structure   Zonal Flow Excitation by Beta-induced Alfv\'en Eigenmode}
\author{Zhiyong Qiu$^{1}$, Liu Chen$^{1,2}$ and Fulvio Zonca$^{3,1}$}

\address{~$^1$Institute for Fusion Theory and Simulation and department of physics, Zhejiang University, Hangzhou, P.R.C. }
\address{~$^2$Dept. of Physics and Astronomy,  Univ. of California, Irvine CA 92697-4575, U.S.A.}
\address{~$^3$ ENEA C. R. Frascati, C. P.
65-00044 Frascati, Italy}

\begin{abstract}
Nonlinear excitation of low frequency zonal structure (LFZS) by beta-induced Alfv\'en eigenmode (BAE) is investigated using nonlinear gyrokinetic theory. It is found that  electrostatic zonal flow (ZF), rather than zonal current, is preferentially excited by finite amplitude BAE. In addition to the well-known meso-scale radial envelope structure, ZF is also found to exhibit  fine radial structure  due to  the localization of BAE with respect to mode rational surfaces. Specifically, the zonal electric field has an even mode structure at the rational surface where radial envelope peaks.
\end{abstract}

\maketitle

\section{Introduction}

Shear Alfv\'en waves (SAW) are expected to play important roles in future magnetic confinement fusion devices  such as ITER. With group velocities mainly along magnetic field lines and frequency close to the characteristic frequencies of fusion alphas, SAWs could be excited by energetic particles (EPs) via wave-particle interactions \cite{YKolesnichenkoVAE1967,SVelikovSPJ1969,AMikhailovskiiSPJ1975,MRosenbluthPRL1975,LChenPoP1994}; and in turn, induce EP transport and degrade overall plasma confinement. This subject has been recently reviewed and discussed in Ref. \citenum{LChenRMP2016}. Among various Alfv\'enic instabilities, beta-induced  Alfv\'en eigenmode (BAE) \cite{WHeidbrinkPRL1993,FZoncaPPCF1996} is of particular  interest since it can be driven unstable by both thermal particles and EPs, with different wavelengths to maximize wave-particle interaction strength. BAE is excited inside the kinetic thermal ion induced SAW continuum gap \cite{LChenNF2007b}, with the frequency in the ion acoustic frequency range and   mode structure highly  localized around magnetic rational surfaces.

There are two routes for the nonlinear saturation of SAWs \cite{LChenPoP2013}, i.e., nonlinear wave-particle and nonlinear wave-wave interactions. Nonlinear  wave-particle interactions of EP driven SAWs, such as BAEs, have been investigated by several different codes, e.g., GTC \cite{HZhangPRL2012} and  HMGC \cite{XWangPRE2012}.  On the other hand, the wave-wave nonlinearity of SAWs in toroidal geometry in the ITER relevant short wavelength regime is relatively less studied.

Among various wave-wave nonlinearities, generation of zonal structures (ZS) by modulational instability is of particular importance. Zonal structures \cite{AHasegawaPoF1979}, including   zonal flows (ZF) and   zonal current (ZC), are known to play important self-regulatory roles on micro-scale drift wave type instabilities by scattering drift waves into short radial wavelength stable domain \cite{LChenPoP2000,ZLinScience1998}.  Typically, the scattering rate depends on the ZS intensity. Thus, understanding the mechanism of ZS generation is of key importance to the nonlinear dynamics of SAWs.  Ref. \citenum{LChenPRL2012} first investigated the nonlinear generation of low frequency zonal structure (LFZS) by Toroidal Alfv\'en eigenmode (TAE), and found that spontaneous excitation of LFZS could be achieved when the  nonlinear drive by TAE overcomes  the threshold condition due to frequency mismatch.  In addition, ZC is found to be  preferentially excited under certain conditions, which are related to the sign of frequency mismatch. These are not those typically verified in tokamak equilibria. In fact, simulations  of TAE nonlinear dynamics,  have observed forced driven of LFZS by TAE rather than spontaneous excitation \cite{ZWangPRL2013,YTodoNF2010}.
Ref. \citenum{HZhangPST2013} studied the nonlinear excitation of ZS by BAE, and found that geodesic acoustic mode (GAM) \cite{NWinsorPoF1968,FZoncaEPL2008} and LFZS are generated in weak and strong EP drive cases, respectively. In the latter case, LFZS is observed to further reduce the final saturation level of BAE.

 In this work, we will investigate the nonlinear excitation of LFZS by BAE using nonlinear gyrokinetic theory \cite{EFriemanPoF1982}  and assuming that BAE initially exists at a prescribed fluctuation amplitude. Self-consistent inclusion of nonlinear wave-particle interactions will be discussed in a later publication.  The rest of the paper is organized as follows. In section \ref{sec:model}, the theoretical model is presented, while   nonlinear LFZS and BAE equations are derived in section \ref{sec:nleq}, which are then used to derive the nonlinear dispersion relation of the modulational instability in section \ref{sec:nldr}. Finally, conclusions and discussions are presented in section \ref{sec:discussion}.

\section{Theoretical Model}\label{sec:model}

We follow the theoretical approach of \cite{LChenPoP2000} and \cite{LChenPRL2012}, and use $\delta\phi$ and $\delta A_{\parallel}$ as the field variables to investigate the nonlinear interactions between BAE and LFZS. To investigate BAE modulational instability,   the fluctuations is assumed to consist of a constant-amplitude pump wave $(\omega_0, \mathbf{k}_0)$ and upper and lower sidebands due to the modulation by LFZS, $(\omega_{\pm},\mathbf{k}_{\pm})$. Using subscripts $Z$ and $B$ for LFZS and BAE, respectively, one then has, $\delta\phi=\delta\phi_Z+\delta\phi_B$, $\delta\phi_B=\delta\phi_0+\delta\phi_++\delta\phi_-$.  Subscripts $0$, $+$ and $-$ denote BAE pump, upper and lower sidebands, respectively. Assuming BAEs have high toroidal mode numbers, we can adopt the well-known ballooning-mode decomposition \cite{JConnorPRL1978} in the $(r,\theta,\phi)$ field-aligned toroidal flux coordinates
\begin{eqnarray}
\delta\phi_0&=&A_0e^{i(n\phi-m_0\theta-\omega_0t)}\sum_j e^{-ij\theta}\Phi_0(x-j)+c.c.,\nonumber\\
\delta\phi_{\pm}&=&A_{\pm}e^{\pm i(n\phi-m_0\theta-\omega_0t)}e^{i(\int \hat{k}_Zdr-\omega_Zt)}\nonumber\\
&&\hspace*{2em}\times\sum_je^{\mp ij\theta}\left\{\Phi_+(x-j) \atop \Phi_-(x-j)\right\}+c.c..\nonumber
\end{eqnarray}
Here, $(m=m_0 +j, n)$ are the poloidal and toroidal mode
numbers, $m_0$ is the reference value of $m$,
$nq(r_0) =m_0$, $q(r)$ is the safety factor, $x=nq-m_0 =
nq'(r-r_0)$,  $\hat{k}_Z\equiv nq'\theta_k$ is the radial envelope wave number in the  ballooning representation, $\Phi$ is the fine radial structure associated with $k_{\parallel}$ and magnetic shear, and $A$ is the envelope amplitude.

Since BAE fine radial mode structure   is highly localized around mode rational surfaces, we   expect that,   LFZS could also exhibit a similar fine radial  structure, in
addition to the well-known meso-scale radial envelope. Thus, we take
\begin{eqnarray}
\delta\phi_Z=A_Ze^{i\int \hat{k}_Zdr-i\omega_Z t}\sum_j\Phi_Z(x-j)+c.c.,
\end{eqnarray}
with $\Phi_Z$ accounting for the fine radial structure.

We assume the same decomposition for $\delta A_{\parallel}$. In the derivation, we   let $\delta\psi=\omega\delta A_{\parallel}/(ck_{\parallel})$ be an alternative field variable  for $n\neq 0$ BAEs, such that ideal MHD condition is recovered if one takes $\delta\phi=\delta\psi$.  Note that  $k_{\parallel}$ as well as $k_r$ should, in general, be considered as  operators.

\section{Nonlinear equations}\label{sec:nleq}

\subsection{Nonlinear LFZS equation}

The first equation of LFZS can be derived from nonlinear vorticity equation \cite{LChenRMP2016}:
\begin{eqnarray}
\noindent&&\frac{e^2}{T_i}\langle(1-J^2_Z)F_0\rangle\delta\phi_Z-\sum_s\left\langle\frac{q}{\omega_Z}J_Z\omega_d\delta H\right\rangle_Z\nonumber\\
&=&-i\frac{c}{B_0\omega_Z}\sum_{\mathbf{k}_Z=\mathbf{k}'+\mathbf{k}''}\hat{\mathbf{b}}\cdot\mathbf{k}''\times\mathbf{k}'\nonumber\\
&&\times\left[\frac{c^2k''^2_{\perp}}{4\pi\omega'\omega''}\partial_l\delta\psi'\partial_l\delta\psi''+\left\langle e(J_kJ_{k'}-J_{k''})\delta \phi_{k'}\delta H_{k''}\right\rangle\right];\nonumber\\
\label{vorticityequation}
\end{eqnarray}
with the nonlinearities coming from    Maxwell  and Reynolds stresses; i.e., the first and second term on the right hand side of equation (\ref{vorticityequation}), respectively  \cite{LChenNF2001}. Here, $J_k=J_0(k_{\perp}\rho)$ with $J_0$ being the Bessel function, $\rho=v_{\perp}/\Omega$, $\Omega$ is the cyclotron frequency, $F_0$ is the equilibrium particle distribution function,  $\sum_s$ is the summation on different species, $q$ is the electric charge,  $\omega_d=(v^2_{\perp}+2
v^2_{\parallel})/(2 \Omega R_0)\left(k_r\sin\theta+k_{\theta}\cos\theta\right)$ is the magnetic drift frequency,  $l$ is the length along the equilibrium magnetic field line; and other notations are standard. Furthermore, $\langle\cdots\rangle$ indicates velocity space integration and $\delta H$ is the nonadiabatic particle response, which can be derived from nonlinear gyrokinetic equation \cite{EFriemanPoF1982}
\begin{eqnarray}
\left(-i\omega+v_{\parallel}\partial_l+i\omega_d\right)\delta H&=&-i\omega\frac{q}{T}F_0J_k\left(\delta\phi+\frac{i}{\omega}v_{\parallel}\partial_l\delta\psi\right)\nonumber\\
&&\hspace*{-10.5em}-\frac{c}{B_0}\sum_{\mathbf{k}=\mathbf{k}'+\mathbf{k}''} \hat{\mathbf{b}}\cdot\mathbf{k}''\times\mathbf{k}'J_{k'}\left(\delta\phi+\frac{i}{\omega}v_{\parallel}\partial_l\delta\psi\right)'\delta H''\label{NLgyrokinetic}.
\end{eqnarray}

Note that the $\mathbf{k}$ are  defined as operators for spatial derivatives; i.e.,
\begin{eqnarray}
\mathbf{k}_0\delta\phi_0&\equiv&[k_{\parallel,0}\mathbf{b}+ k_{\theta,0}\hat{\mathbf{\theta}}-inq'\partial_x \ln\Phi_{0}\hat{\mathbf{r}}]\delta\phi_{0},\nonumber\\
\mathbf{k}_{\pm}\delta\phi_{\pm}&\equiv&[\pm k_{\parallel,0}\mathbf{b}\pm k_{\theta,0}\hat{\mathbf{\theta}}+(\hat{k}_Z-inq'\partial_x \ln\Phi_{\pm})\hat{\mathbf{r}}]\delta\phi_{\pm},\nonumber\\
\mathbf{k}_Z\delta\phi_Z&\equiv&(\hat{k}_Z-inq'\partial_x \ln\Phi_Z)\hat{\mathbf{r}}\delta\phi_Z.\nonumber
\end{eqnarray}
In the rest of the paper, for the simplicity of notations,   the subscript $``0"$ will be suppressed when appropriate.

Noting $|k_{\perp}\rho_i|\ll1$ in the inertial layer, one then obtains from surface averaged vorticity equation
\begin{eqnarray}
-i\omega_Z\chi_{iZ}\delta\phi_Z&=&-\frac{c}{2B_0}k_{\theta}\frac{\partial^2}{\partial r^2}\left[\delta\phi_0\partial_r\delta\phi_-
-\delta\phi_-\partial_r\delta\phi_0\right.\nonumber\\
&&\left.+\delta\phi_+\partial_r\delta\phi_{0^*}-\delta\phi_{0^*}\partial_r\delta\phi_+\right.\nonumber\\
&&-(k^2_{\parallel}V^2_A/\omega^2_0)\left(\delta\psi_-\partial_r\delta\psi_0-\delta\psi_0\partial_r\delta\psi_-\right.\nonumber\\
&&\left.\left.-\delta\psi_+\partial_r\delta\psi_{0^*}+ \delta\psi_{0^*}\partial_r\delta\psi_+\right)\right].\label{LZFSVorticity0}
\end{eqnarray}
Here, $V_A=B_0/\sqrt{4\pi n m_i}$ is the Alfv\'en speed,  $\chi_{iZ}$ is the well-known neoclassical polarizability of LFZS \cite{MRosenbluthPRL1998}, and we have  $\chi_{iZ}\simeq 1.6 k^2_Z\rho^2_i/\sqrt{\epsilon}$ with $\epsilon\equiv r/R_0$ being the inverse aspect ratio of the torus.
Applying the $\delta\phi_B$ representation into equation (\ref{LZFSVorticity0}), we then obtain,
\begin{eqnarray}
-i\omega_Z\hat{\chi}_{iZ}A_Z\Phi_Z&=&-(c/2B_0)\hat{k}_Zk_{\theta}\left(1-(k^2_{\parallel}V^2_A/\omega^2_0)\right)\nonumber\\
&&\times\left(A_0A_--A_{0^*}A_+\right)\left|\Phi_0\right|^2.\label{LFZSVorticity}
\end{eqnarray}
Here, $\hat{\chi}_{iZ}\equiv\chi_{iZ}/(k^2_Z\rho^2_i)\simeq1.6q^2/\sqrt{\epsilon}$ \cite{MRosenbluthPRL1998}. In deriving equation (\ref{LFZSVorticity}), we assumed $\Phi_+\simeq\Phi_0$, $\Phi_-\simeq\Phi^*_0$ (self-consistently proved {\sl a posteriori}), and $\Phi_0$ being purely real. Note that $\Phi_0$ may have anti-Hermitian part if the pump BAE is driven via wave-particle resonances; and, in that case, the nonlinear term derived here should be modified to take into account the contribution of resonant particles  \footnote{Here, we take $\delta\phi_0\partial_r\delta\phi_--\delta\phi_-\partial_r\delta\phi_0$ as an example. $\delta\phi_0\partial_r\delta\phi_--\delta\phi_-\partial_r\delta\phi_0=A_0A_-\exp(i\int \hat{k}_Zdr-i\omega_Zt)\sum_{l,p}\exp(-i(l-p)\theta)\left(i\hat{k}_Z\Phi_0\Phi_-+\Phi_0\partial_r\Phi_--\Phi_-\partial_r\Phi_0\right)$. If $\Phi_-\simeq\Phi^*_0$ and $\Phi_0$ has an anti-hermitian part, we have $\Phi_0\partial_r\Phi_--\Phi_-\partial_r\Phi_0\simeq-2i \mathbb I{\rm m}(nq'\partial_x\ln\Phi_0)|\Phi_0|^2$, which could dominate over $\hat{k}_Z|\Phi_0|^2.$}.

Note that $A_Z$ is the usual ``meso-scale" radial envelope, while $\Phi_Z$ corresponds to the fine radial structure of ZF. Taking advantage of the scale separation, it is clear  from equation (\ref{LFZSVorticity}) that  we can let
 \begin{eqnarray}
\Phi_Z=|\Phi_0|^2.
\end{eqnarray}
The envelope equation for $A_Z$ is then given by \cite{FZoncaPoP2004}
\begin{eqnarray}
\omega_Z\hat{\chi}_ZA_Z&=&-i\frac{c}{2B_0}k_{\theta}\hat{k}_Z\hat{C}\left(A_0A_--A_{0^*}A_+\right).\label{ZFenvelope}
\end{eqnarray}
Here, $\hat{C}\equiv \langle\langle1- k^2_{\parallel}(x)V^2_A /\omega^2_0\rangle\rangle$, with  $\langle\langle\cdots\rangle\rangle$ defined as
\begin{eqnarray}
\langle\langle\cdots\rangle\rangle\equiv \int \left(\cdots\right)|\Phi_0|^2(x) dx\left/\int |\Phi_0|^2 dx,\right.\nonumber
\end{eqnarray}
and we typically have $\hat{C}\simeq 1$ for modes near the accumulation point of BAE gap \cite{FZoncaPPCF1996} \footnote{The correction to $\hat{C}$ from $k_{\parallel}\neq0$ poloidal harmonics is of order $O(q^4\beta^2\epsilon^2)$ smaller.}.

Noting $\partial_rA_{\pm}=\pm i\hat{k}_ZA_{\pm}$,   equation (\ref{LFZSVorticity}) can also be straightforwardly expressed as
\begin{eqnarray}
A_Z\Phi_Z=\frac{1}{2}\frac{c}{B_0}k_{\theta}\frac{\hat{C}}{\omega_Z\hat{\chi}_{iZ}}\left(A_0\partial_rA_--A_{0^*}\partial_rA_+\right)|\Phi_0|^2.
\end{eqnarray}
Furthermore, with  $|\hat{k}_Z|\ll|\partial_r\ln\Phi_Z|\sim |nq'|$, we then have
\begin{eqnarray}
\delta E_r&=&-\frac{c\hat{C}k_{\theta}}{2B_0\omega_Z\hat{\chi}_{iZ}}\left(A_0\partial_rA_--A_{0^*}\partial_rA_+\right)\partial_r|\Phi_0|^2.\nonumber\\
\end{eqnarray}
Thus, assuming $A_{\pm}\sim \cos (\hat{k}_Z r)$, we then have $\delta E_r$ as an even function of $r$ at the center of the envelope;  consistent with the GTC simulation \cite{HZhangPST2013} \footnote{Note that in the GTC simulation \cite{HZhangPST2013}, BAE is driven by energetic particles (EP), while EP effect is not considered here. Inclusion of EPs could change the parity of the radial electric field here.}.

The other  equation of ZF can be derived from the parallel component of the nonlinear ideal Ohm's law.
\begin{eqnarray}
 \delta E_{\parallel,Z} &=-&\sum_{\mathbf{k}'+\mathbf{k}''=\mathbf{k}_Z}\hat{\mathbf{b}}\cdot\delta \mathbf{u}_{k'}\times\delta\mathbf{B}_{k''}/c
\end{eqnarray}
with $\delta\mathbf{u}$ being the $\mathbf{E}\times\mathbf{B}$ drift velocity. Noting $k_{\parallel,Z}=0$, one then has
\begin{eqnarray}
&&\frac{\partial\delta A_{\parallel, Z}}{\partial t}=-i\frac{c}{B_0}k_{\theta}\frac{\partial}{\partial r}\left(\frac{k_{\parallel,0}}{\omega_0}\delta\psi_0\delta\phi_0-\frac{k_{\parallel,-}}{\omega_-}\delta\psi_-\delta\phi_0\right.\nonumber\\
&&\hspace*{2em}\left.+\frac{k_{\parallel,+}}{\omega_+}\delta\psi_+\delta\phi_{0^*} -\frac{k_{\parallel,0^*}}{\omega_{0^*}}\delta\psi_{0^*}\delta\phi_+\right)\nonumber\\
&&=-\frac{c}{B_0}k_{\theta}\frac{1}{\omega^2_0}\frac{\partial}{\partial t}\frac{\partial}{\partial r}\left(k_{\parallel,0}\delta\phi_0\delta\phi_-+k_{\parallel,0}\delta\phi_+\delta\phi_{0^*}\right).\label{LFZSOhm1}
\end{eqnarray}
In deriving equation (\ref{LFZSOhm1}), we have applied ideal MHD condition ($\delta\phi-\delta\psi=0$) for BAEs in the inertial layer \cite{FZoncaPPCF1996},   $k_{\parallel,\pm}=\pm k_{\parallel,0}$, and $\omega_{\pm}=\pm\omega_0+i\partial_t$. We then obtain
\begin{eqnarray}
\delta A_{\parallel,Z}=-\frac{c}{B_0\omega^2_0}k_{\theta}\frac{\partial}{\partial r}(k_{\parallel}\delta\phi_0\delta\phi_-+k_{\parallel}\delta\phi_+\delta\phi_{0^*}).\label{LFZSOhm}
\end{eqnarray}

\subsection{Nonlinear BAE equations}

To derive the nonlinear dispersion relation of the parametric process, we need also the equations describing BAE sidebands generation.
The first nonlinear equation of BAE sidebands can be derived from nonlinear  Ohm's law.  Noting $\delta E_{\parallel}=-\partial_l(\delta\phi-\delta\psi)$,
we then have,
\begin{eqnarray}
(\delta\phi-\delta\psi)_{\pm}&=&\frac{1}{B_0} k_{\theta}\left\{\delta\phi_0 \atop \delta\phi_{0^*}\right\}\left(\frac{c}{\omega_0}\partial_r\delta\phi_Z-\frac{1}{k_{\parallel}}\partial_r\delta A_{\parallel,Z}\right).\nonumber\\
\label{BAEOhm}
\end{eqnarray}

To close the system, we derive the other equation of BAE sidebands from nonlinear vorticity equation. Here, we give the detailed derivation for BAE upper sideband.   Following \cite{FZoncaPoP2004}, and applying $k_{\perp}\rho_i\ll1$ in the inertial layer, one obtains
\begin{eqnarray}
&&\frac{c^2}{4\pi\omega^2_+}B_0\frac{\partial}{\partial l}\frac{k^2_{\perp}}{B_0}\frac{\partial}{\partial l}\delta\psi_++\frac{e^2}{T_i}\langle(1-J^2_k)F_0\rangle\delta\phi_+\nonumber\\
&-&\sum\left\langle\frac{q}{\omega}J_k\omega_d\delta H\right\rangle_+\nonumber\\
&=&-\frac{c}{B_0\omega_+}k_{\theta}\frac{n_0e^2}{T_i}\frac{\rho^2_i}{2}\left[\partial_r\delta\phi_Z\partial^2_{\perp}\delta\phi_0-\delta\phi_0\partial^3_r\delta\phi_Z \right.\nonumber\\
&-&\left.\frac{V^2_A}{\omega^2_0}\left(k_{\parallel}\delta\psi_0\partial^3_r(k_{\parallel}\delta\psi_Z)-\partial_r(k_{\parallel}\delta\psi_Z)\partial^2_{\perp}(k_{\parallel}\delta\psi_0)\right)\right].\nonumber\\ \label{BAEVorticity1}
\end{eqnarray}
Here, $\delta\psi_Z\equiv \omega_0\delta A_{\parallel,Z}/(ck_{\parallel,0})$. Noting that $\langle\langle k^2_{\parallel,0}V^2_A/\omega^2_0\rangle\rangle\ll1$, the second term on the right hand side is negligible comparing to the first term. Thus, because of finite plasma compressibility, the Alfv\'enic state is broken \cite{LChenPoP2013,LChenRMP2016}, and Maxwell's stress does not nearly cancel Reynold's stress as for incompressible SAWs in uniform plasma. Meanwhile, we have that
\begin{eqnarray}
&&\partial_r\delta\phi_Z\partial^2_{\perp}\delta\phi_0-\delta\phi_0\partial^3_r\delta\phi_Z\nonumber\\
&=&\left[-i\hat{k}_Z(k^2_{\theta}-\hat{k}^2_Z)+nq'(-k^2_{\theta}\partial_x\ln\Phi_Z+3\hat{k}^2_Z\partial_x\ln\Phi_Z)\right.\nonumber\\
&+&\left.i\hat{k}_Z(nq')^2\left((\partial_x\ln\Phi_0)^2-3(\partial_x\ln\Phi_Z)^2\right)\right.\nonumber\\
&+&\left.(nq')^3\left(\partial_x\ln\Phi_Z(\partial_x\ln\Phi_0)^2-(\partial_x\ln\Phi_Z)^3\right)\right]\delta\phi_0\delta\phi_Z. \nonumber\\
\end{eqnarray}
Noting that the BAE envelope equation is derived by averaging out the fine structures, and  that $\Phi_Z$ and $\Phi_0$ are even functions of $x$ in the vicinity of rational surfaces, only the first (proportional to $-i\hat{k}_Z(k^2_{\theta}-\hat{k}^2_Z)$) and third (proportional to $(nq')^2$) terms will contribute; while the terms proportional to $nq'$ and $(nq')^3$ are odd functions of $x$.  Substituting equation (\ref{BAEOhm}) into equation (\ref{BAEVorticity1}), and noting $\langle\langle k^2_{\parallel}V^2_A/\omega^2_0\rangle\rangle\ll1$, one then obtains
\begin{eqnarray}
k^2_{\perp,+}L_{B,+}A_+\Phi_+&=&i\frac{c}{B_0\omega_+}k_{\theta}\hat{k}_Z\left[k^2_{\theta}-\hat{k}^2_Z\right.\nonumber\\
&&\hspace*{-10em}\left.-(nq')^2\left((\partial_x\ln\Phi_0)^2-3(\partial_x\ln\Phi_Z)^2\right)\right]A_ZA_0\Phi_0\Phi_Z.\label{VorticityUp1}
\end{eqnarray}
Here, $L_B\equiv1-k^2_{\parallel}V^2_A/\omega^2-\omega^2_G/\omega^2$ is the WKB dispersion relation of BAE, with $\omega_G$ being the accumulation point frequency of kinetic thermal ion SAW continuum gap (``BAE gap"). It is easy to see that  the second term on the RHS of equation (\ref{VorticityUp1}), due to the fine structure, is   larger than the first term by $O(\hat{k}^2_Z/(nq')^2)\gg1$. Keeping     terms from fine radial structure only, assuming $\Phi_0=\exp(-x^2/(2\Delta^2_r))/(\pi^{1/4}\Delta^{1/2}_r)$ with $\Delta_r\ll1$ being the characteristic scale length of the fine radial structure, and defining $\mathscr{E}_B\equiv\int \Phi^*_0L_B\Phi_0 dx$, equation (\ref{VorticityUp1}) can then be solved perturbatively. Expanding $L_{B,+}=L^{(0)}_{B,+}+L^{(1)}_{B,+}+\cdots$ with $L^{(0)}_{B,+}=L_{B,0}$, $\mathscr{E}_+=\mathscr{E}^{(0)}_++\mathscr{E}^{(1)}_++\cdots$, $\Phi_+=\Phi_0+\Phi_1+\cdots$ with $\Phi_1$ orthogonal to $\Phi_0$, we then have, to the lowest order
\begin{eqnarray}
\mathscr{E}^{(0)}_+=\mathscr{E}_0\equiv\int \Phi^*_0 L_{B,0}\Phi_0 dx.
\end{eqnarray}
To the next order, we obtain the required nonlinear equation for BAE upper sideband
\begin{eqnarray}
k^2_{\perp,+}\mathscr{E}^{(2)}_+A_+=i\frac{11}{2\sqrt{2}\sqrt{\pi}\Delta^3_r}\frac{c}{B_0\omega_+}k_{\theta}\hat{k}_ZA_ZA_0.\label{VorticityUp}
\end{eqnarray}
Here,  $\mathscr{E}^{(2)}_+\equiv \int \Phi^*_0 L^{(2)}_{B,0}\Phi_0 dx$.

The nonlinear equation for BAE lower sideband can be derived similarly
\begin{eqnarray}
k^2_{\perp,-}\mathscr{E}^{(2)}_-A_-=-i\frac{11}{4\sqrt{\pi}\Delta^3_r}\frac{c}{B_0\omega_-}k_{\theta}\hat{k}_ZA_ZA_{0^*} .\label{VorticityLow}
\end{eqnarray}

\section{Nonlinear dispersion relation}\label{sec:nldr}

The nonlinear dispersion relation describing LFZS generation by BAE  can then be derived. Substituting equations (\ref{VorticityUp}) and (\ref{VorticityLow}) into equation (\ref{LFZSVorticity}), we then obtain
\begin{eqnarray}
\omega_Z\hat{\chi}_{iZ}&=&-\frac{1}{2}\left(\frac{c}{B}k_{\theta}\hat{k}_Z\right)^2\hat{\xi}|A_0|^2\nonumber\\
&&\times\left(\frac{1}{k^2_{\perp,+}\mathscr{E}^{(2)}_+\omega_+}+\frac{1}{k^2_{\perp,-}\mathscr{E}^{(2)}_-\omega_-}\right),
\end{eqnarray}
with $\hat{\xi}\equiv 11/(2\sqrt{2\pi}\Delta^3_r)$. Taking $\omega_+\simeq\omega_0$, $\omega_-\simeq-\omega_0$, $\mathscr{E}^{(2)}_{\pm}\simeq(\partial\mathscr{E}_0/\partial\omega_0)(\pm\omega_Z+\Delta)$, with $\Delta\equiv\hat{k}^2_Z (\partial^2\mathscr{E}_0/\partial \hat{k}^2_Z)/(2\partial\mathscr{E}_0/\partial\omega_0)$ being the frequency mismatch, and taking $\gamma\equiv-i\omega_Z$, we finally derive the  following modulational instability  dispersion relation
\begin{eqnarray}
\gamma^2=-\Delta^2+\left(\frac{c}{B}k_{\theta}\hat{k}_Z\right)^2|A_0|^2\frac{\hat{\xi}}{k^2_{\perp,+}\hat{\chi}_{iZ}\omega_0(\partial\mathscr{E}_0/\partial\omega_0)}.
\end{eqnarray}
Thus,  LFZS can be excited when the nonlinear drive overcomes the threshold condition due to frequency mismatch. We note that  the nonlinear coupling effects for the parametric process would be underestimated by one order of magnitude if the contribution from fine structure was ignored.  The threshold condition can then be estimated as
\begin{eqnarray}
\left|\frac{\delta B_r}{B_0}\right|_{threshold}\sim O(10^{-4}).
\end{eqnarray}
In estimating this threshold condition, we assumed $\Delta\sim \sqrt{\beta} \omega_0$, $q^2\sim O(10)$, $k^2_{\perp}\rho^2_i\sim O(10^{-1})$ and $\rho_i/R_0\sim O(10^{-3})$. Here $\beta$ is the ratio of  thermal and magnetic pressure.

\section{Conclusions and discussions}\label{sec:discussion}

In conclusion, we have derived the  equations describing the nonlinear interactions between low frequency zonal structure (LFZS) and beta-induced Alfv\'en eigenmode (BAE), which are then used to derive the nonlinear dispersion relation of the modulational instability. It is found that, predominantly  electrostatic zonal flow (ZF) can be spontaneously excited by a finite-amplitude BAE when the threshold condition due to frequency mismatch is exceeded.

We also found that  the obtained scalar potential of LFZS, which is related to the electrostatic ZF, has both the usual meso-scale radial envelope  as well as a fine radial structure due to the short scale  of BAE localized near the mode rational surface. The inclusion of fine structure  significantly increases the nonlinear coupling coefficients  of the modulational interactions, and thereby, decreases the threshold condition on BAE amplitude.  The corresponding radial electric field, meanwhile, is an even function with respect to  the mode rational surface where envelope peaks; consistent with the observations in GTC simulation \cite{HZhangPST2013}.

Due to    $\langle\langle k^2_{\parallel}V^2_A/\omega^2_0\rangle\rangle\ll1$, electron nonlinear dynamics described by Maxwell stress is much smaller than Reynolds stress, such that the ``pure Alfv\'enic state" is strongly broken by finite ion compressibility here. As a consequence, the breaking of   ideal MHD condition (i.e., equation (\ref{BAEOhm})) does not affect the final nonlinear dispersion relation, contrary to the TAE cases  \cite{LChenPRL2012,ZQiuEPL2013} (e.g., equations (11), (12) and (13) of Ref. \citenum{ZQiuEPL2013}). Furthermore, zonal current is also much weaker compared with ZF; again in contrast to the TAE case. However, we note that, if  BAE was excited by energetic particles localized away from rational surfaces, $k_{\parallel}$ could be finite, leading to finite ZC generation.

Further inclusion of contributions due to resonant particles; such as energetic particles, in  the BAE nonlinear dynamics is of crucial importance. The saturation of BAE can be due to the competition/collaboration of the wave-wave nonlinearity, discussed here, and wave-particle nonlinear dynamics \cite{XWangPRE2012,HZhangPRL2012}.  In fact, even if one ignores the nonlinear evolution of the driving resonant particles   \cite{FZoncaNJP2015}, the linear response related to wave-particle resonances may lead to important modifications to the  picture discussed here, as noted in the specific comments and remarks in our analysis. These additional effects associated with the presence of resonant particles will be further explored in future studies.

\section*{Acknowledgments}
This work is supported by     US DoE GRANT,  the ITER-CN under Grants Nos.  2013GB104004  and   2013GB111004,  the National Science Foundation of China under grant Nos.  11575157  and 11235009, Fundamental Research Fund for Chinese Central Universities and   EUROfusion Consortium
under grant agreement No. 633053.

\end{document}